\documentclass[12pt]{iopart}

\usepackage{iopams}  
\usepackage{amsmath}
\usepackage[colorlinks, citecolor = blue, urlcolor = blue]{hyperref}
\usepackage{graphicx}
\begin{document}

\title[Field tuning Kitaev systems for spin fractionalization and topological order]{Field tuning Kitaev systems for spin fractionalization and topological order}

\author{ J Das$^1$, S Kundu$^2$,
A Kumar$^3$ and V Tripathi$^1$}
\address{$^1$ Department of Theoretical Physics, Tata Institute of
Fundamental Research,  Mumbai,  India}
\address{$^2$ Department of Physics, University of Florida, Gainesville, Florida 32611, USA}
\address{$^3$ National High Magnetic Field Laboratory, Tallahassee, Florida, 32310, USA}
\ead{vtripathi@theory.tifr.res.in}

\begin{abstract}
The honeycomb Kitaev model describes a $Z_2$ spin liquid with topological order and fractionalized excitations consisting of gapped $\pi$-fluxes and free Majorana fermions. Competing interactions, even when not very strong, are known to destabilize the Kitaev spin liquid. Magnetic fields are a convenient parameter for tuning between different phases of the Kitaev systems, and have even been investigated for potentially counteracting the effects of other destabilizing interactions leading to a revival of the topological phase. Here we review the progress in understanding the effects of magnetic fields on some of the perturbed Kitaev systems, particularly on fractionalization and topological order.   
\end{abstract}

%
%
%
%
%

\section{\label{sec:introduction}Introduction}

Quantum spin liquids (QSLs) are among the most interesting paramagnetic quantum phases of matter with long range entanglement and fractionalized emergent excitations \cite{anderson1987resonating,baskaran1993resonating,kalmeyer1987equivalence,kivelson1987topology,affleck1988large, read1991large, wen1989vacuum,wegner1971duality,senthil2000z,kogut1979introduction,moessner2001resonating,kitaev2006anyons,savary2016quantum}. The honeycomb Kitaev model, which happens to be integrable, describes a quantum spin liquid where the lattice spins are fractionalized into free Majoranas fermions subjected to a static $Z_2$ gauge field \cite{kitaev2006anyons}. The model also has $\pi$-flux excitations or visons (Wilson loops), which are gapped. These properties make it a deconfined phase. More importantly, the ground state of the Kitaev model has $Z_2$ topological order (same as the Toric code) which means it is four-fold degenerate on the torus, the excitations are abelian semionic anyons, and the topological entanglement entropy $\gamma$ is nontrivial, with a value $\gamma=\ln 2 .$ The Majorana excitations in the Kitaev model are gapped or gapless depending upon the relative strength of the interactions in the three bonds emanating from each site, with the gapless majoranas appearing when the magnitudes of the three interaction strengths satisfy the triangle inequality. Applying a small Zeeman field in certain directions (such that the magnetic field has nonzero components along the three spin projection axes) gaps out the Majorana modes in the gapless phase and makes them chiral. A half-quantized thermal Hall conductivity is expected in this regime, and attributed to chiral gapless majoranas on the edge. Such Majorana edge modes leading to half-quantized thermal Hall response are also expected for some topologically trivial states such as chiral $p$-wave superconductors \cite{ivanov2001non}. However in the Kitaev model, this low-field chiral phase also has a nonabelian Ising topological order (ITO) characterized by intrinsic nonabelian anyonic excitations in the bulk, a three-fold degeneracy of the ground state on the torus, and a finite topological entanglement entropy $\gamma=\ln 2.$ Both the Toric code and nonabelian ITO phases of the Kitaev model are useful from the point of view of topological quantum computing - especially in the nonabelian phase, qubit operations are realized by braiding sub-gap Majorana excitations. Spin fractionalization and topological order are thus the two most remarkable features of the Kitaev model. 

Considerable attention has been devoted to Zeeman effects in Kitaev systems. The non-commutation of Zeeman perturbation with the Kitaev Hamiltonian induces $\pi$-flux (vison) fluctuations. In his original work, Kitaev studied the effect of weak magnetic fields in the gapless regime ignoring vison fluctuations by projecting the Zeeman perturbed Hamiltonian to the flux free sector associated with the ground state of the Kitaev model. In the effective model thus obtained, Zeeman perturbations with nonzero field components along the three natural spin quantization axes induce a three-spin interaction that breaks time reversal and parity symmetries (similar to three spin scalar spin chirality terms in spin rotation invariant models) but is otherwise diagonal in the flux sector. The Majorana excitations also remain eigenstates of the model; however, they now become gapped and chiral. Kitaev demonstrated that this phase is associated with a half-quantized thermal Hall conductivity. Later it was shown that the Zeeman field also imparts a dispersion to the visons, and their contribution to properties such as the thermal Hall response has attracted attention recently. The behaviour at higher fields is dictated by the sign of the Kitaev interaction $K.$ For ferromagnetic (FM) sign of the Kitaev coupling, the system directly transitions into a topologically trivial partially spin-polarized phase. The transition to the topologically trivial phase appears to be a confinement-deconfinement type driven by vison proliferation. In the antiferromagnetic (AFM) case, there is at least one intermediate field phase that separates the low field ITO phase and the high field partially spin-polarized paramagnetic phase. The nature of the intermediate field phase in the AFM Kitaev model is still not well understood and there is ongoing debate even about a basic matter as to whether it is a topologically trivial phase or not. At high fields, spin waves (and not the fractionalized Kitaev quasiparticles) are expected to provide a good approximation for the many-body excitations. Topological order in the ferromagnetic case is fragile, and is easily degraded by, say, a $(111)$ Zeeman field of the order of $10^{-2}\times K,$ while for the antiferromagnetic case, the field required (for the same orientation) is around ten times larger. 

Apart from Zeeman perturbations, various competing spin interactions (e.g. Heisenberg-type exchange, anisotropic and off-diagonal interactions) also induce vison fluctuations and degrade topological order, often resulting in the appearance of magnetic order. Motivated by Kitaev materials, a significant effort has already gone into understanding different magnetic phases in the parameter space of such interactions, and the likelihood that for a given set of parameters, how close a magnetically ordered Kitaev material is to the Kitaev spin liquid phase \cite{janssen2019heisenberg,hermanns2018physics,takagi2019concept,trebst2022kitaev}. Changes in the symmetry and order of the ground state are reflected in the quasiparticle excitations, which are essentially created through making certain local moves on the ground state. Quasiparticle character is expected to evolve from the fractionalized Majoranas and visons in the Kitaev spin liquid phase to spin waves in the magnetically ordered or high field phases. The parameter regimes straddling these extremes hold the promise of hosting both conventional and fractionalized quasiparticles. For example, in the magnetically ordered phases very close to the Kitaev spin liquid, the excitations can have a substantial Kitaev-like character that ultimately goes away as one moves deeper into the magnetically ordered phase \cite{kumar2020kitaev}. Such magnetically ordered phases, in which energy windows exist where quasiparticles have characteristics of the adjacent spin liquid phase have been termed as proximate spin liquid (PSL) phases.  
While Zeeman fields and non-Kitaev spin interactions both degrade the topological order, not much is known about their interplay when they are simultaneously present. One possibility is that the two kinds of perturbations likely have very different and competing mechanisms for degrading topological order, which tempts one to use them simultaneously in a manner that neither is able to suppress the topological order existing nearby in parameter space. Another possibility is that both degrading mechanisms not only compete with each other, but also act in tandem to destroy topological order. In the latter situation, using field suppression of, say, magnetic order would not prove to be a viable strategy for resurrecting the lost topological order.

Disorder is ultimately unavoidable, and thus, it becomes important to understand the effects they have on the physical properties of the Kitaev system and the role they play in the degradation of topological order. When disorder is weak, it may even serve as a way to detect the underlying topological phase. For example, site vacancies in the Kitaev model are known to host localized nonabelian anyons and these fractionalized magnetically active states are known to impart specific singularities to the field and temperature dependence of impurity susceptibility. Disorder signatures in other phases in Kitaev systems, brought about through field tuning or competing interactions, are currently not well understood.

From the experimental perspective, material realization of this phase has proved elusive. One of the main obstructions is the presence of competing exchange interactions that may be small compared to the Kitaev interaction, but nevertheless significant enough to cause a long range magnetic ordering of the spins. Indeed, some of the most celebrated Kitaev materials fall in this category. Both $\alpha$-RuCl$_3$ and Na$_2$IrO$_3$ - first generation Kitaev materials with ferromagnetic Kitaev interaction - have antiferromagnetic order which is zigzag type in the Kitaev planes, but fully three-dimensional. However the large ratio of the Curie-Weiss temperature (order of the Kitaev scale) to antiferromagnetic (AFM) ordering temperature suggests strong frustration effects and proximity to the spin liquid phase. The magnetic order can be suppressed either thermally or through Zeeman fields. This naturally raises the question if some of the excitations in the magnetically disordered phases thus obtained have a substantial resemblance to the fractionalized Kitaev quasiparticles, and whether any signatures of topological order could be detected here. New generation Kitaev materials are designed to be more two-dimensional which frustrates the tendency for 3D magnetic ordering. Both FM and AFM Kitaev interactions have been found, and the latter looks very promising given the greater robustness of topological order in AFM Kitaev systems. Unfortunately, enhancing two-dimensionality comes at a cost because the abundant interplanar space allows strong interstitial and substitutional disorder which is another parameter that degrades topological order. The resulting states are often glassy and there is some evidence for underlying fractionalized excitations. Not only the strongly disordered spin liquids are insufficiently understood, but they are also less likely to see a revival of topological order through parametric tuning compared to their disorder free or low disorder counterparts. Consequently, here we will mostly focus on systems that are not very strongly disordered. We will review available experimental data on some of the Kitaev materials covering field-dependent transport (e.g. thermal conductivity, thermal Hall conductivity), magnetometry (high field torque response, susceptibility) and inelastic neutron scattering for clues. 

This paper reviews the current experimental and theoretical understanding of magnetic field effects on the fractionalization and topological order of the honeycomb Kitaev system. The models we are interested in have the general form
\begin{equation}
H =  - K \sum_{\langle ij \rangle \in \gamma \text{-links}} S_{i}^{\gamma}S_{j}^{\gamma} + \sum_{i}\mathbf{h}\cdot\mathbf{S} + H',
   \label{eq:basic-model}
\end{equation}
where $H'$ represents other competing spin interactions. While modeling material systems, the components of the vectors are taken along the three spin projection axes of the system and they do not normally coincide with the crystallographic axes. In his original work \cite{kitaev2006anyons}, Kitaev studied the effect of small Zeeman perturbations in the flux-free sector and showed that in the gapless Majorana phase of the Kitaev model, such perturbations introduce a gap, 
\begin{equation}
 \Delta_M \sim \frac{|h_x h_y h_z|}{{\Delta_V}^2},
 \label{eq:majorana-gap}
\end{equation}
in the Majorana spectrum. Here $\Delta_V$ is the flux or vison gap, which for the ferromagnetic Kitaev model is $\Delta_V \sim 10^{-2} K.$ Strictly speaking, the validity of Eq. \ref{eq:majorana-gap} is limited to $|\mathbf{h}| < \Delta_{V}$ within the ITO phase, and higher fields degrade the topological order through vison fluctuations.

 This review is structured as follows.  Sec. \ref{sec:expt} is devoted to  the review of experiments on Kitaev materials for signs of fractionalization and/or topological order, especially involving magnetic fields. Particularly in Subsec. \ref{sec:scattering} we demonstrate  the results of  scattering and spectroscopic probes for fractionalization e.g. inelastic neutron scattering \cite{banerjee2016proximate} vs interacting magnons \cite{winter2016challenges}, Raman and terahertz spectroscopy,  NMR and NQR probes. Subsec. \ref{sec:thermal-expt}  is regarding  the thermodynamic response in Zeeman fields, where  we have described specific heat/thermal conductivity and thermal Hall response - field dependence (including direction), high field torque magnetometry, etc. In Sec. \ref{sec:theory}, we briefly explain the  review of field-tuned phases in Kitaev and perturbed Kitaev systems - both numerical and analytical studies. Subsec. \ref{sec:pure-kitaev} shows the Zeeman effects in the pure Kitaev limit  whereas Subsec. \ref{sec:competing-int} is about field effects with competing spin interactions.  The theoretical study of response in Kitaev systems is summarised in Subsec. \ref{sec:theory-responsel}.   We end this review with some discussion and future directions in Sec. \ref{sec:discuss}.
\section{\label{sec:expt} Experimental scene}
Kitaev materials have been a subject of extensive investigation for identifying possible parameter tuning into regimes where ITO and fractionalized excitations could exist. Among these materials, $\alpha$-RuCl$_3$ stands out. While it orders antiferromagnetically (zigzag) at $T_N \approx 7 K,$ the ordering temperature is much smaller than the Curie-Weiss temperature of $\Theta_{CW} \approx -150K$, indicating magnetic frustration. The other workhorse Kitaev material has been the alkali iridate Na$_2$IrO$_3,$ also with zigzag order, and $T_N \approx 15 K,$ $\Theta_{CW} \approx -130K.$ Experimentally, some of the techniques that are commonly adopted for
exploring possible signatures of fractionalized excitations in Kitaev
magnets include (a) spectroscopic probes such as inelastic neutron scattering (INS), Raman scattering and time-dependent terahertz spectroscopy, NMR and NQR, and (b) thermodynamic probes. Below we review some of the key experiments that help shed light on this question.
\begin{figure}
\centering
\includegraphics[width=\columnwidth]{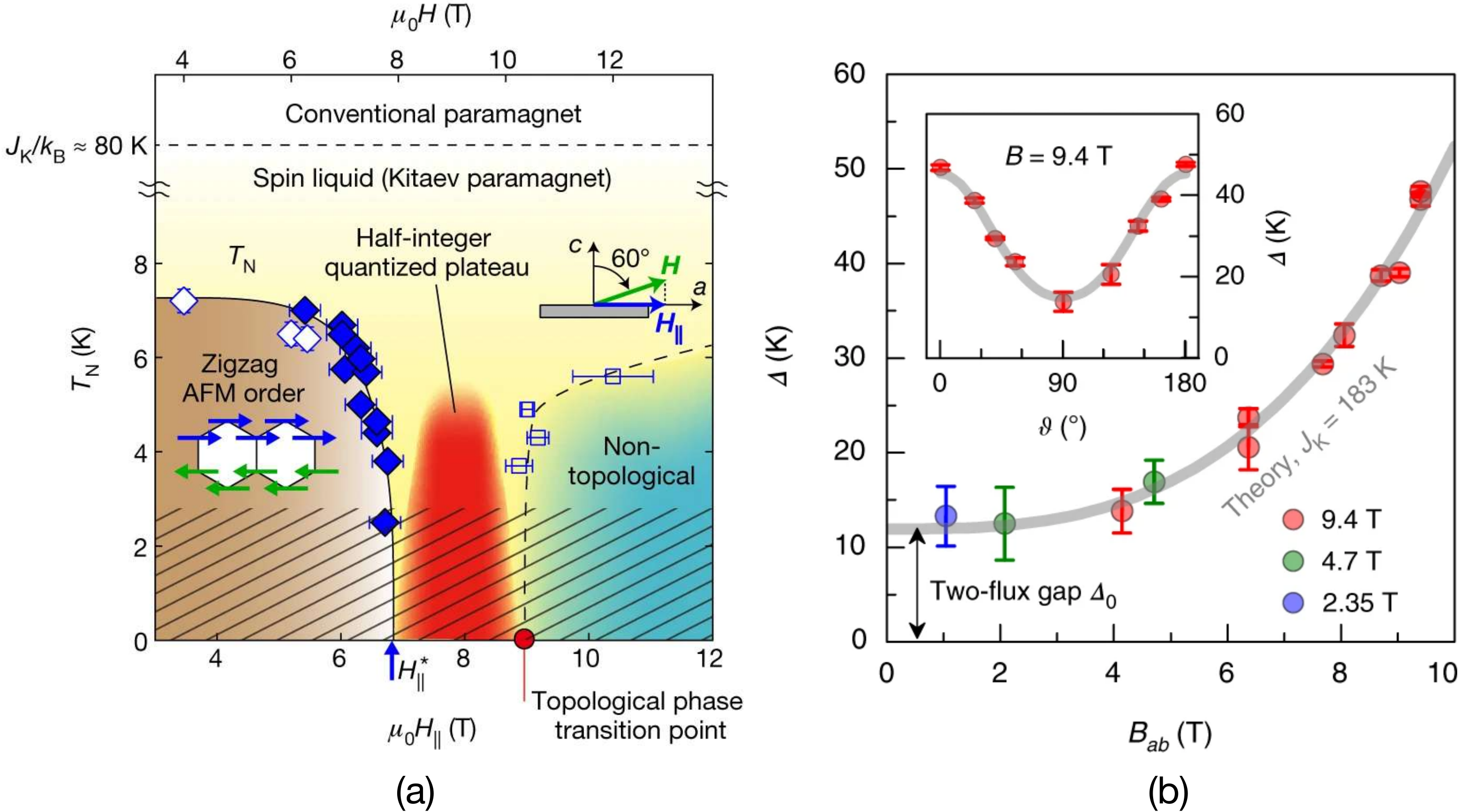}
\caption{\label{fig:hall-gap} In (a) the phase diagram
of $\alpha$-RuCl$_3$ in tilted field is shown as described in Ref \cite{kasahara2018majorana}, indicating the field  revived KSL phase from AFM Zigzag order. (b) (Ref. \cite{janvsa2018observation}) describes   the field-dependent spin excitation gap which is the sum of two terms: the two-flux gap $\Delta_0$ and the gap acquired by Majorana fermions in a weak magnetic field, theoretically predicted to be proportional to the cube of the field. The inset shows $\Delta (\theta)$ for B = 9.4 T where B is applied at an angle $\theta$ from the $ab$ plane. (Reprinted with permission from Ref. \cite{kasahara2018majorana} and Ref. \cite{janvsa2018observation})}
\end{figure}
\subsection{\label{sec:scattering}Scattering and spectroscopic probes for fractionalization}
\textit{Inelastic neutron scattering}: In INS measurements on $\alpha$-RuCl$_3$, the high-energy part of a persistent broad continuum
centered at the $\Gamma-$point, that remains essentially unchanged
until high temperatures of the order of 100 K, has been interpreted
as a signature of fractionalized excitations. In zero field, the momentum
dependence of the scattering was seen to resemble the response
calculated for the AFM Kitaev QSL at zero temperature, whereas the
constant Q response at the $\Gamma-$point was seen to be closer to that calculated
for a FM Kitaev model in an effective magnetic field \cite{banerjee2018excitations}. Since the current understanding is that the sign of the Kitaev interaction in this material is definitely ferromagnetic (FM), we believe that the fits to the momentum dependence of scattering obtained from AFM Kitaev models are not very relevant here. On the other hand, a ferromagnetic Kitaev with additional antiferromagnetic $\Gamma$ type anisotropic interaction also gives a good match with the INS data \cite{ran2017spin,wang2017theoretical}. It is not yet known whether fractionalized excitations can appear through field tuning of such a $K$-$\Gamma$ model. 

INS measurements were also carried out with
a finer field step, higher energy resolution and greater field strength \cite{zhao2022neutron}. For an in-plane field, spin-wave
excitations around the $M-$point are gradually suppressed and vanish
around the critical field of 7.5 T, while a continuum observed near
the $\Gamma-$point under zero field still persists when the spin
waves vanish. Based on the gap evolution of the continuum, it is concluded
that the intermediate-field spin liquid phase is possibly gapless. 
In a related study, Balz et al. \cite{balz2019finite} explored $\alpha$-RuCl$_3$ under external magnetic fields using INS, and found the absence of well-defined magnons near field-suppressed magnetic order, suggesting a magnetically disordered state. 
In another work \cite{winter2017breakdown}, it has been argued that the presence of a broad excitation continuum and the absence of clear magnon peaks does not prove the existence of fractionalized excitations, and such broadening of the magnons could also originate from magnon-magnon interactions. 

Inelastic neutron scattering measurements data have thus been interpreted in both ways - either as consistent with the existence of fractionalized excitations \cite{banerjee2016proximate} or as a manifestation of magnon-magnon interactions \cite{winter2017breakdown,winter2016challenges}. Clearly, a positive test of fractionalization is desirable, instead of 
arguments that merely indicate consistency with fractionalization.
 
\textit{Raman spectroscopy}: A magnetic continuum with spin-wave excitations has been inferred from Raman measurements \cite{wulferding2020magnon} at low fields, suggesting simultaneous presence of excitations with spin-wave and possibly fractionalized character. At higher fields, the spectral weight is transferred to well-defined sharp excitations corresponding to one-magnon and multimagnon bound states, while at intermediate fields, a weakly bound state emerges which does not connect smoothly to them. This mode has been argued by the authors of Ref. \cite{wulferding2020magnon} as consisting of bound states of itinerant Majorana fermions, since flux excitations are largely invisible to the Raman scattering process . The magnetic field-induced QSL phase was also addressed via helicity-dependent Raman scattering, which could potentially capture anyonic excitations that are chiral \cite{sahasrabudhe2023chiral}. The field regime between 7.5-10.5 T is characterized by clear spectroscopic signatures, such as a plateau of the Raman optical activity of a dominant, chiral spin-flip excitation. However the chiral excitations could also be topological magnons, the quasiparticles expected in the Kitaev model at higher magnetic fields in the partially polarized paramagnetic phase. 

\textit{Terahertz spectroscopy}: Time-dependent terahertz spectroscopy (TDTS) probes the continuum spectrum with high sensitivity and energy resolution, yielding an absolute measurement of the imaginary part of the dynamical magnetic susceptibility at zero wave vector. The THz spectra determined for parallel and perpendicular orientation of the static and THz magnetic fields, up to 7 T, showed \cite{wu2018field} two sharp resonances at 2.5 and 3.2 meV and broader features in the range 4-6 meV that appear only at applied fields above approximately 4 T. The authors of Ref. \cite{wu2018field} interpreted this behaviour using linear spin wave theory, taking into account a $C_{3}-$breaking distortion of the honeycomb lattice and the presence of structural domains.

\textit{NMR and NQR probes}: It has also been shown using NMR \cite{janvsa2018observation} that the field-dependence of the spin excitation gap can be fitted by the sum of a finite zero-field value and a cubic growth in the low-field region (see Fig. \ref{fig:hall-gap}(b)). Since a single spin flip is expected to create a pair of visons and a Majorana, the low field behaviour provides a positive test for the existence of fractionalized Kitaev excitations. The spin excitations of $\alpha-$RuCl$_{3}$
have also been investigated through combined NMR and NQR spectroscopy for temperatures
down to 0.4 K, well below the magnetic ordering temperature. Upon field-suppression of magnetic order, two kinds of gapped excitations are observed, and interpreted as evidence for Majorana hybridization by the external field in a re-emergent ITO phase \cite{nagai2020two}. However, since spin-wave analysis in the partially polarized high field phase would also yield two gapped modes, this latter observation does not by itself resolve the issue.

 \subsection{\label{sec:thermal-expt} Thermodynamic response in Zeeman fields}
 Thermodynamic response is a valuable resource for probing fractionalization in Mott insulator spin liquid systems. The dramatic observation in 2010 \cite{yamashita2010highly} of a linear-$T$ thermal conductivity in a certain organic triangle lattice antiferromagnetic insulating spin liquid material has been regarded as a key evidence for a fractionalized (spinon) Fermi surface. Thermal measurements have also been widely employed for probing possible fractionalization and topological order in the Kitaev materials, especially $\alpha$-RuCl$_3.$ 
 
 \textit{Thermal Hall response}: The most important claim in this context has been the observation of half-quantized thermal Hall conductivity $\kappa_{xy}$ in the vicinity of field-suppressed magnetic order \cite{kasahara2018majorana, yokoi2021half, kasahara2022quantized,bruin2022robustness}. An in-plane magnetic field of strength $h \lesssim 10 \text{T}$ is sufficient to degrade  the magnetic order and the half-quantized thermal Hall response was reported for a field window $7 \text{T} \le h\le 9 \text{T},$ with one report at even higher fields \cite{kasahara2018majorana} (see Fig.\ref{fig:hall-gap}(a) for details).  
 However other groups have strongly contested these claims \cite{kasahara2018majorana}, contending  that while the thermal Hall response in the vicinity of field-suppressed magnetic order is indeed large (often significantly exceeding half-quantization), at low temperatures $\kappa_{xy}/T$ appears to fall sharply below the $\pi k_{B}^{2}/12\hbar$ value expected for half-quantization, suggesting the absence of gapless chiral edge modes. These studies have either supported interpretations based on chiral magnons (see also Sec. \ref{sec:theory}) or even phonon scattering from spin-dependent excitations or disorder. Proponents of the phonon mechanism point out that the thermal Hall response is seen even along directions normal to the Kitaev-planes \cite{lefranccois2023oscillations} and that its temperature and field evolution tracks the longitudinal thermal conductivity that is expected to be phonon-dominated. We have a word of caution here, for phonons will in general couple to magnons since lattice vibrations, through their effect on electron-electron interactions, result in local electronic spin excitations. This is more likely in frustrated magnets where multiple competing states are found in close proximity. Thus the phonon excitations are not quite decoupled from the magnetic ones. The same arguments also lead to the expectation that such magneto-phonon excitations should be susceptible to decay to fractionalized emergent excitations. Sans a microscopic calculation of thermal Hall response or a more direct experimental test, it would not be correct to rule out any of these physical mechanisms. Later in the review we will discuss the theoretical treatments for the thermal Hall response of perturbed Kitaev systems.

\textit{Thermal conductivity}: Along with the sizable thermal Hall effect, $\alpha$-RuCl$_3$ also displays unusual features in the longitudinal thermal conductivity.  Multiple “oscillations” of the longitudinal thermal conductivity have been observed over a range of fields around field-suppressed magnetic order. These are not quantum oscillations \cite{lefranccois2023oscillations,czajka2021oscillations} because the fields in question are much smaller than the values requires for inserting $O(1)$ flux quanta in the unit cells. A possible origin, suggested in the literature \cite{lefranccois2023oscillations,czajka2021oscillations}, is a succession of field-induced canted magnetically ordered states. 

\textit{Specific heat}: In contrast to thermal Hall measurements that have hitherto not been able to establish field-induced ITO, a clearer evidence of fractionalization is available \cite{tanaka2022thermodynamic} from specific heat measurements in the vicinity of field-suppressed magnetic order in $\alpha$-RuCl$_3.$  The authors of Ref. \cite{tanaka2022thermodynamic} observed that for the field $\mathbf{h}\parallel \mathbf{a}$ -- a configuration that has nonzero components along all three spin projection axes -- the field dependence of the excitation gap extracted from the specific heat data is consistent with the behaviour in Eq. \ref{eq:majorana-gap}. However, when the field is in the direction $\mathbf{h}\parallel \mathbf{b},$ which corresponds to $h_x h_y h_z =0,$ the specific heat reportedly has a $T^2$ behaviour expected for 2D fermionic Dirac dispersion (or even 2D magnons). Such interpretation of the specific heat data is based on a crucial assumption that the excitations in the magnetically disordered phase obtained through field induced suppression of magnetic order in $\alpha$-RuCl$_3$ are similar in nature to the fractionalized Majorana excitations of the Kitaev model.

Over a larger range of temperatures exceeding the Curie-Weiss scale, one of the most common
features observed is the characteristic two-peak structure in the
heat capacity, after properly accounting for the phonon background,
corresponding to the entropy release of localized fluxes and itinerant
Majorana fermions. The Kitaev model has a characteristic two-peak
spectrum of the heat capacity, separated in temperature by at least
one order of magnitude. Experimentally, a clear double-peak structure
is observed at all the magnetic fields investigated, and the high-temperature
anomaly remains almost constant as a function of external fields,
and is indicative of itinerant Majorana fermions \cite{widmann2019thermodynamic}. 

\textit{Other thermometry probes}: From magnetocaloric, thermal expansion and magnetostriction data in
$\alpha-$RuCl$_{3}$ single crystals, it was shown that an apparent
energy gap structure that evolves when the low-temperature antiferromagnetic
order is suppressed, can show a cubic field dependence and remain
finite at zero field, depending on how the thermal expansion data
are modeled \cite{schonemann2020thermal}.

\textit{Torque magnetometry measurements:} 
The appearance of nonmonotonic behavior and a peak-dip feature in
the torque response, around magnetic fields corresponding to the scale of the zigzag ordering temperature, in both the Kitaev
materials Na$_{2}$IrO$_{3}$ \cite{das2019magnetic}
and $\alpha-$RuCl$_{3}$ \cite{leahy2017anomalous}, has been interpreted
as evidence for a field-induced phase transition from the zigzag ordered
state to a state with no magnetic order and no simple spin polarization.
The size of the anomaly in the torque response is found to drop significantly
for temperatures exceeding the magnetic ordering temperature, 
indicating that this behavior is connected to the presence of magnetic
order. In the case of Na$_{2}$IrO$_{3}$, the observed signature
in the torque response has been used to constrain the effective spin
models for these classes of Kitaev materials to ones with dominant
FM Kitaev interactions, while excluding alternative models with dominant
AFM Kitaev interactions. At high magnetic fields, the
long-range spin correlation functions have been shown to decay rapidly,
pointing towards the possibility of a transition to a field-induced QSL. More recently, the presence of the intermediate QSL phase
in $\alpha-$RuCl$_{3}$ has been a subject of debate,
with certain experiments indicating that a moderate in-plane field
of about 7 T may induce an intermediate QSL phase before the polarized
phase \cite{wellm2018signatures,kasahara2018majorana,balz2019finite,yokoi2021half,nagai2020two} and other experimental evidence, such as that based on magnetization,
magnetocaloric and torque measurements, pointing towards a single transition with no such intermediate phase \cite{johnson2015monoclinic,kubota2015successive,bachus2020thermodynamic}. Recent theoretical work
has suggested the absence of the intermediate QSL phase under in-plane
fields and predicted its presence for out-of-plane fields, with
two phase transitions involved in the process. Notably, there has
been a detailed study of the magnetization of $\alpha-$RuCl$_{3}$
for fields in various directions within the honeycomb plane and along
the $c^{*}-$axis perpendicular to it, up to about 100 T \cite{zhou2023possible}. Under fields applied along and close to the $c^{*}-$axis,
an intermediate phase was found, bounded by two transition fields
of about 32.5 T and 83 T. 
For a tilt of about $9^{\circ}$ from the $c^{*}-$axis, only the
lower transition field was observed to be present, which decreased for larger tilt
angles, while the intermediate phase disappeared. These results were
supported by DMRG calculations 
on the $K-J-\Gamma-\Gamma^{\prime}$ model. 

\section{\label{sec:theory} Theoretical studies}

We will review the theoretical progress in the following order. First we will examine the literature devoted to the pure Kitaev model subjected to Zeeman fields focusing on aspects such as fractionalization, topological order and signatures. Thereafter we will review the current understanding of competing spin interaction effects on similar aspects. Properties that one can experimentally measure to seek answers to the above questions will also be discussed. Here we would pay particular attention to the theory of the thermal Hall effect since, as we argue below, some of the approaches in wide use that are based on the spin wave approximation are not correct.

\subsection{\label{sec:pure-kitaev}Zeeman effects in the Kitaev limit}

The low-energy excitations of the isotropic Kitaev model consist entirely of free, gapless Majorana excitations and no visons. Kitaev showed \cite{kitaev2006anyons} that in the presence of a small Zeeman perturbation, the model, upon projecting to the vison-free ground state sector, results in an effective low-field Hamiltonian containing three-spin perturbations, 
\begin{equation}
H_{\text{low}} =  - K \sum_{\langle ij \rangle \in \gamma \text{-links}} S_{i}^{\gamma}S_{j}^{\gamma} + \kappa \sum_{\langle ijk\rangle }S_{i}^{\alpha}S_{j}^{\beta}S_{k}^{\delta} + \text{geometrical equivalents},
   \label{eq:effective-model}
\end{equation}
that breaks parity and time reversal symmetry and is equivalent to scalar spin chirality type perturbations in models of Heisenberg magnets. Here $\kappa \propto h_x h_y h_z/K^{2},$ the sites in the triad $\langle ijk \rangle$ are nearest neighbours, and the components $\alpha,$ $\beta,$ $\delta$ are such that $ij\in \alpha\text{-link},$ $jk\in \delta\text{-link},$ and $\beta\neq \alpha,\,\delta.$ The above model is integrable and has the same excitations as the unperturbed Kitaev model (static, gapped visons and free Majorana fermions). If the field has nonzero components along all the three natural spin quantization axes, a nonzero Majorana gap $\Delta_M \approx |h_x h_y h_z|/\Delta_{V}^{2}$ appears with characteristic cubic scaling with the field strength, a peculiar direction dependence, and inversely proportional to the square of the vison gap $\Delta_V.$ For the FM Kitaev model, $\Delta_V \approx 0.025 K.$ The Majorana fermions in this gapped phase are chiral, and while the bulk is gapped, the boundary hosts a gapless edge mode, and given that the filling fraction of the edge fermions is $1/2,$ which Kitaev argued  will result in a half-quantized thermal Hall response. These arguments relied on the noninteracting mean field model. Kitaev did not however demonstrate using any bulk-edge correspondence that the edge modes are associated with a conformal field theory (CFT) with a chiral central charge $c_{-}=\nu/2,$ where $\nu=\pm 1$ is the Majorana Chern number. By considering half-vortex twist defects in the bulk (not intrinsic excitations of the mean field theory), he argued that the vortex cores would host zero energy sub-gap majoranas that obey non-abelian braiding rules of Ising topological order (ITO). Here he made the observation that the mean field model is equivalent to a chiral p-wave superconductor - a phase that has no topological order \cite{ivanov2001non} but is known to allow Majorana zero modes in half-vortices that satisfy ITO braiding rules. Assuming that $c_{-}=\nu/2$ is correct, Kitaev reasoned that this would be consistent with a bulk ITO phase although he had demonstrated ITO braiding rules not for intrinsic excitations but for half-vortices.
\begin{figure}
\centering
\includegraphics[width=\columnwidth]{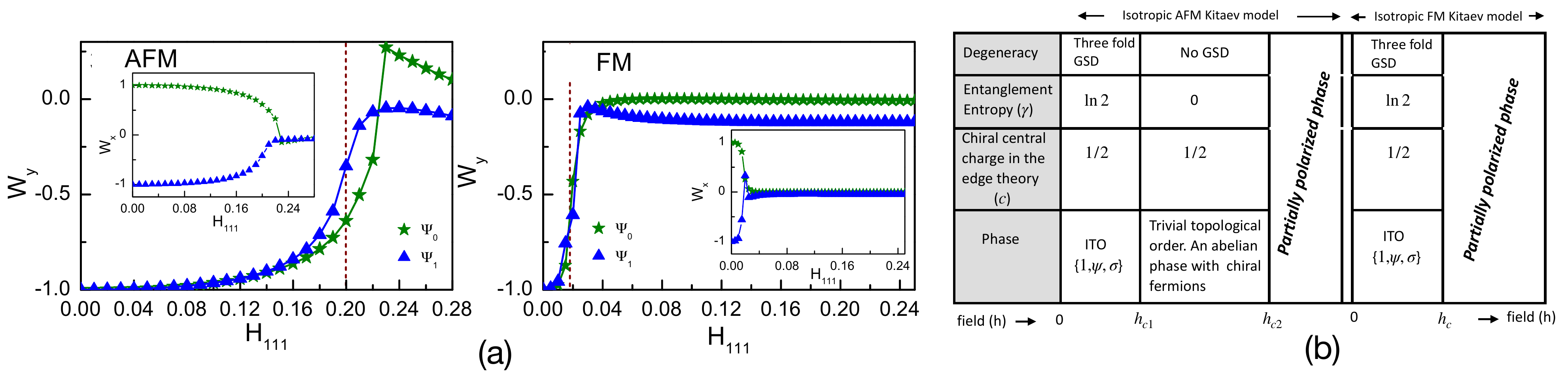}
\caption{\label{fig:table-wilson}: In (a) (Reprinted with permission from Ref. \cite{zhu2018robust}), the expectation values of  Wilson loop operators $W_y$ (main panel) and  $W_x$ (inset) for the two lowest energy states for both AFM  and FM  models are shown as a function of $(111)$ Zeeman field. The two states have $W_y=-1$ but are distinguished by $W_x=\pm 1$. (b) (from Ref. \cite{das2024microscopic}) describes all the topological properties (ground state degeneracy, topological entropy, quasi-particles) and chiral central charge for both FM and AFM Kitaev models in different $(001)$ field  regimes.}
\end{figure}

Subsequently, a number of works provided additional evidence in support of Kitaev's ITO proposition. One way to probe the topological order numerically, using exact diagonalization, has been to study the expectation of fluxes (Wilson loops) associated with non-contractible loops on the torus \cite{zhu2018robust} (see Fig. \ref{fig:table-wilson}(a)), which were found to take definite values $\pm 1$ for one of the loops and only $-1$ for the other. However because of finite size effects, the three-fold ground state degeneracy expected of ITO could not be established. In another work \cite{gohlke2018dynamical}, density matrix renormalization group (DMRG) techniques were used to calculate the von Neumann bipartite entanglement entropy for open cylinder geometries. The entropy was shown to have an ``area law'' component and a constant ``topological'' component, where the latter had values $\gamma = \ln2,$ or $\gamma = (1/2)\ln2,$ both of which were argued to be consistent with ITO and the difference was on account of boundary conditions. Unlike the mean field theories, the DMRG calculation accounts for both the fermion as well as gauge contributions. Zeeman induced vison fluctuations, although small in the ITO phase, are nevertheless interesting, since the erstwhile static visons now acquire a dispersion. For the FM Kitaev model, the dispersing visons are not topological \cite{joy2022dynamics,chen2023nature} and have a trivial non-projective translational symmetry \cite{chen2023nature}. In contrast, for the AFM case, the visons have a nonzero Chern number of $\pm 1$ \cite{joy2022dynamics,chen2023nature} and a nontrivial projective translational symmetry with $\pi$-flux per unit cell.   

Larger Zeeman fields are known to destabilize  the ITO in both AFM and FM Kitaev model. However for a given strength of the Kitaev coupling, the topological order is more robust in case of AFM model compared to the FM case. For the AFM case, there is an ongoing debate surrounding the existence and nature of an intermediate abelian phase between the low field ITO and high field partially polarised paramagnetic phase whereas the in the FM model, one directly transits from the ITO to the partially polarised phase.
From DMRG and ED calculations \cite{zhu2018robust,patel2019magnetic}, the transition takes place  in the pure Kitaev model with a $(111)$ oriented Zeeman field at $h \approx 0.02K$ for the FM Kitaev coupling and the intermediate regime persists in a field interval $0.2K \leq h \leq 0.36K $  in the AFM case. Subsequently a debate has arisen about the nature of the intermediate phase for Zeeman $(111)$ fields, for example with the authors of Ref. \cite{patel2019magnetic,jiang2018field,hickey2019emergence,hickey2021generic} contending that it was a gapless $U(1)$ spin liquid with a spinon fermi surface, while in Ref. \cite{zhang2022theory}, using a combination of parton mean field theory and numerics, it was argued that the intermediate field is gapped and chiral (with $\nu=\pm 4.$) Another work \cite{jiang2020tuning} identifies a second narrow region with fractionalized abelian excitations and ($\nu=\pm 4$) followed by a wider topologically trivial phase with $\nu =\pm 1.$ As many as five different field tuned phases have been reported in recent DMRG studies of the Kitaev model on a 1D ladder setting, with an emergent glassy phase appearing at intermediate fields \cite{yogendra2023emergent}. 
In Ref. \cite{gohlke2018dynamical}, the spin-diagonal zero momentum response $S^{xx}(0,\omega)$ was studied using tensor network techniques for the AFM Kitaev model in a $(111)$ Zeeman perturbation together with the scalar spin chirality terms. Small magnetic fields tend to reduce the Majorana and flux gaps and also give additional features in the dynamical response function that the authors interpret as flux dispersion. At intermediate fields, the zero momentum gap becomes much smaller (possibly gapless within numerical resolution), and at large fields, a magnon mode grows out of the continuum. The features in the intermediate phase could not be properly interpreted whether they arose from fractionalized excitations or from magnon broadening through magnon-magnon interactions. Approaching from the high field end, a more recent work \cite{schellenberger2022dynamic} has examined the one and two quasiparticle (dressed magnon) features in the dynamical structure factor, supporting the picture that from the high field side, the broadening seen at lower fields is on account of magnon interactions. Clearly, the intermediate phase for $(111)$ field orientation is not fully understood.  

In contrast, there is a better understanding for $(001)$ Zeeman perturbations and we confine our attention to the AFM case where there is an intermediate phase \cite{nasu2018successive,berke2020field}. A parton mean field study of the Kitaev model with a $(001)$ field shows a gapless spectrum persisting up to $h \approx 0.5 K;$ however this gapless regime actually consists of two phases separated by $h_{C1}\approx 0.42K$ and $h_{C2}\approx 0.5K$ with the low-lying wavefunctions in the two regimes having opposite chiralities \cite{nasu2018successive}. At higher fields, a partially polarized trivial phase appears. Remarkably, there is a dimensional reduction of the excitation spectrum in the high field partially polarized phase from 2D to 1D dispersing spin waves along the $xy$-chains \cite{nasu2018successive,feng2023dimensional,das2023jordan}. The authors of Ref. \cite{feng2023dimensional} further found from numerical calculations that the intermediate gapless phase consists of weakly-coupled 1D quantum Ising critical fermionic chains described by a $(1+1)$D CFT with a central charge $c=1/2.$ Upon introducing a small scalar spin chirality, the spectrum becomes gapped \cite{nasu2018successive} and chiral with $\nu=\pm 1,$, and the parton Chern number is found to change sign at $h_{C1}.$ Based on the change of sign of the parton Chern number from the ITO to the intermediate phase, the authors of Ref. \cite{nasu2018successive} offered that this would manifest in a change of sign of the half-quantized thermal Hall conductivity. 
We have already discussed above that mean field parton theory is not sufficient for a complete characterization of a spin liquid, and it is essential to take the contribution from the gauge sector into account. Recently two of us \cite{das2024microscopic} microscopically derived and analyzed an effective abelian Chern-Simons gauge theory using a Jordan-Wigner fermionization strategy. For this, the authors of Ref. \cite{das2024microscopic} considered gauge fluctuation effects around the mean field theory of Ref. \cite{nasu2018successive}. They have  provided a comprehensive characterization of the low-field  ITO  by studying GSD, $\gamma,$ fusion rules and braiding statistics of the quasi-particles. Unlike Kitaev’s case, their non-abelian anyons are intrinsic bulk excitations and is one of the few instances where a non-abelian phase may
be described within an abelian gauge theory framework. Despite the change of sign of the Chern number at $h_{C1},$ it was found that upon taking the contribution from the gauge sector into account, the chiral central charge remains unchanged at $c_{-} = 1/2$, implying no change of sign of the half quantized thermal Hall response. Although the AFM  intermediate phase has topologically trivial order  with no fractionalized excitations, a unique ground state, and no topological entanglement entropy; nevertheless it is not a trivial product state and belongs to the class of a $p+ip$ superconductor where fermion number parity is not gauged. The complete results are summarized in Table \ref{fig:table-wilson} (b).

\subsection{\label{sec:competing-int}Zeeman effects with competing spin interactions}

There is already a vast literature on the possible phases of Kitaev models subjected to competing spin interactions and magnetic fields (see e.g. Ref. \cite{janssen2019heisenberg,hermanns2018physics,takagi2019concept,trebst2022kitaev}). However we are interested in a more restricted question, namely the signatures of fractionalization and topological order and changes to such properties upon field tuning. Such questions should not be viewed in isolation - for example, in the context of magnetically ordered underdoped cuprates, there has been an effort to understand the properties (especially an anomalously large thermal Hall response) originating from fractionalized (semionic) excitations within the magnetically ordered phase \cite{samajdar2019enhanced}. In our opinion a similar situation is indicated in the characteristic cubic field scaling of the spin gap from NMR measurements in the Kitaev material $\alpha$-RuCl$_3$ for small Zeeman perturbation within the magnetically ordered phase \cite{samajdar2019enhanced}.      

\textit{Microscopic models - phenomenology and numerics}: Perhaps the simplest models with competing interactions that one can study are those with $K-\Gamma$ and $K-J$ type of interactions, together with a Zeeman field. Although these toy models are not immediately relevant for the Kitaev materials, they provide a simple example of how the competition of two phases, one of which is a Kitaev spin liquid, is affected by the action of a magnetic field. Ref \cite{liu2018dirac} reports a study of $K-\Gamma$ model on the honeycomb lattice in an external magnetic field by parton mean field and variational Monte Carlo calculations. The partons move in a mean field that corresponds to zigzag magnetic order - however the zigzag order here appears to be imposed only as a parameter and not the result of a phase competition. Depending on the field orientation, the authors predict the existence of a field induced Dirac spin liquid (e.g. for $(001)$ Zeeman fields) or an abelian Kalmeyer-Laughlin type chiral spin liquid with $\nu=\pm 2,$ where the latter would show an integer quantized thermal Hall effect. The analysis does not take into account the gauge fluctuation effects required to impose $U(1)$ charge conservation and the understanding of the nature of the spin liquid phases and their excitations could change once the gauge fluctuations are taken into account. In Ref. \cite{lee2020magnetic}, possible field induced quantum phases in theoretical models of the Kitaev magnets (Ferromagnetic Kitaev - antiferromagnetic $\Gamma$) is studied, using the two dimensional tensor network approach or infinite tensor product states. Various magnetically ordered and paramagnetic states are reported apart from the expected chiral Kitaev spin liquid occupying a small area in the phase diagram (i.e. small $\Gamma/K,\,h/K$). The finite field paramagnetic phases of interest have nematic order. The nematic phases in the $K-\Gamma$ model give away to the complex magnetic orders with large unit cells when $\Gamma/K$ becomes large. Neither of these two papers shed light on the nature of topological order or fractionalized excitations, if any, in the field induced paramagnetic phases. More complicated models that better describe the parent Kitaev materials have also been investigated. In a recent study \cite{li2021identification} a number of numerical techniques were deployed to understand possible field induced  spin liquid phases in a $(K-J-\Gamma-\Gamma')$ model with FM Kitaev interactions, with parameters chosen from fits to certain established experimental results on $\alpha$-RuCl$_3.$ Finite-temperature calculations suggested that in $\alpha$-RuCl$_3,$ a window of excitation energies above the magnetization gap exists and is possibly a Kitaev phase (similar to proposals in some INS experiments) but the nature of the paramagnetic phase was not proved explicitly. When the magnetic field is applied perpendicular to the honeycomb plane, the zigzag order is suppressed at 35 T, above which, and below a polarization field of 100 T level, there emerges a field-induced paramagnetic phase distinct from a partially polarized phase that exists at higher fields. While ED and DMRG calculations suggest the intermediate field phase to be gapless, the VMC calculations suggest an abelian gapped chiral spin liquid phase with $\nu = \pm 2.$ Intermediate field gapless phases have also been found in Ref.  \cite{yu2023nematic}  for fields in the $(111)$ direction, where spin correlations were found to decay as a power law in distance and the specific heat obeyed a $T^{2}$ law. Note that experimentally, a gapped phase is indicated at intermediate fields in specific heat measurements \cite{janvsa2018observation} except for certain directions where the Majorana gap in the Kitaev model vanishes. In summary, the existing studies of Kitaev models perturbed by various spin interactions and subjected to a magnetic field have not reported any evidence for field revived ITO although there is some support for intermediate paramagnetic phases that could be regarded as abelian spin liquids. 
Furthermore, in the analytical studies, gauge effects have not been taken into account, which limits the confidence with which one can claim anything about the nature of topological order and fractionalized excitations. 

\textit{Gauge theory phenomenology}: A deeper analysis of the possibility of field-induced ITO in magnetically ordered Kitaev systems has been made recently using a phenomenological gauge theory approach. For example in \cite{zou2020field}, the field tuned  criticality in $\alpha$-RuCl$_3$ has been described using models of critical bosons or gapless fermions coupled to emergent non-abelian QCD$_3$ Chern-Simons gauge theory. Although not microscopically derived, these models are designed to give a field-theoretical description of a pre-determined sequence of phases, namely zigzag (ZZ) AFM order $\rightarrow$ ITO $\rightarrow$ partially polarized paramagnet. The ITO phase is proposed to consist of an $SU(2)_2$ CS theory and a $U(1)_{-4}$ CS theory (both of which have Majorana excitations) followed by hybridizing the Majorana fermions in these two theories (i.e. an anyon condensation). The zigzag AFM to ITO and ITO to polarized transitions are respectively described by $N_f = 1$ and $N_f=2$ QCD$_3$ theories. Since there is no independent microscopic support for the above sequence of transitions, it remains an open question whether this gauge theory is suitable for the Kitaev materials where field-tuned re-emergence of ITO is being investigated.

\subsection{\label{sec:theory-responsel}Response in Kitaev systems}
A number of proposals have been made theoretically for the experimental detection of fractionalization in Kitaev materials using magnetic field tuning. These studies, like we reviewed earlier for experiments, can be broadly categorized as spectroscopic or thermodynamic probes. 

\subsection*{\label{sec:theory-spectroscopy}Spectroscopic probes of Zeeman perturbed Kitaev models} Apart from the dynamical spin structure factor (governing INS) studies that we have already described above, we mention here some more studies of response in the Kitaev limit. Ref. \cite{nasu2019nonequilibrium} used a time dependent mean field approach to study the ``transient Fermi surface'' of the Majorana excitations created following a $(001)$ Zeeman quench from a finite value to zero at some instant of time. In particular, for the AFM case, such transient Fermi surfaces were argued to be topologically different from those at low fields. More recently, it was shown \cite{harada2023field} that it is possible to control the local vison density by spatio-temporal variation of a $(001)$ field. 

\subsection*{\label{sec:theory-thermalhall} Thermal response}
Thermal Hall conductivity $\kappa_{xy}$ is the leading thermal transport probe for Kitaev systems, with the latter providing a direct test of the topological order. The low temperature thermal Hall response of a gapped chiral state is determined by the chiral central charge $c_{-}$ of the edge excitations. For the ITO phase, $c_{-} = 1/2,$ which translates to a $1/2$-quantized thermal Hall conductivity. This value of $c_{-}$ is not unique to ITO and also appears, for example, in a topologically trivial phase such as a $p+ip$ superconductor (see our earlier discussion). Thus, half-quantized thermal Hall response is a necessary but not sufficient condition for establishing ITO.

The chiral central charge can in principle be obtained provided one has the knowledge of the low-energy effective theory in the relevant parameter regime. As we have seen above, for realistic models of Kitaev materials, we do not yet have a good understanding of the field induced paramagnetic phases realized upon suppression of the magnetic order present in the parent material. At high fields, in the partially polarized phase, spin waves provide a good description of the low-lying physics.

Linear response calculation of thermal Hall response is complicated on two counts. First, unlike its electrical counterpart where one has a TKNN \cite{thouless1982quantized} formula relating the Chern number of the bulk ground state to the Hall conductivity, thermal Hall conductivity must be calculated for geometries with a boundary. Second, the linear response thermal Hall conductivity contains two important components \cite{cooper1997thermoelectric} - a Kubo part which is a correlator of energy currents, and an energy magnetization part which represents bound currents that must be subtracted from the local energy current since they do not contribute to the transport thermal Hall current. In practice, it is difficult to obtain the Kubo and energy magnetization contributions in interacting systems, and almost all the past studies have been conducted for models with free fermion \cite{qin2011energy} or bosonic excitations \cite{matsumoto2014thermal}. More recently there has been progress in the finite temperature thermal Hall response of interacting fermionic matter \cite{guo2020gauge} where the contribution of gauge fluctuations has been addressed in a large-$N$ theory. Another approach has been to consider the contribution of vison dispersion in the ITO phase to the thermal Hall response \cite{joy2022dynamics,chen2023nature}. While the half-integer quantization does not appear to be affected at such low fields, $\kappa_{xy}$ displays a nonmonotonous temperature dependence in the AFM Kitaev model where the visons are in general chiral.    

For noninteracting fermions, taking into account both the Kubo and energy magnetization contributions yields the following expression for the thermal Hall conductivity (see Ref. \cite{qin2011energy}):
\begin{align}
    \kappa^{\text{tr}}_{xy}=-\frac{1}{e^2 T_0}\int d\epsilon (\epsilon-\mu_0)^2 \Theta_{xy}(\epsilon)\epsilon f'(\epsilon),
\end{align}
where $\Theta_{xy}(\epsilon)=\frac{2e^2}{\hbar}\sum_{\epsilon_{nk}\leq \epsilon} \text{Im}<\frac{\partial u_{nk}}{\partial {k_x}}|\frac{\partial u_{nk}}{\partial {k_y}}>$ is the total Berry curvature of the fermionic bands up to an energy $\epsilon.$

In magnetically ordered or partially polarized paramagnetic phases of the perturbed Kitaev systems, the spin wave approximation is a reasonable starting point. 
In Ref. \cite{matsumoto2014thermal}, the following expression for the bulk thermal Hall response of noninteracting bosons was obtained ($\hbar=k_B = 1$),
\begin{equation}
    \kappa_{xy}=-\frac{T}{V}\sum_{k}\sum_{n=1}\Big[c_2(n_{B}(\xi_{nk}))-\frac{\pi^2}{3} \Big] \Omega_{nk},
    \label{eq:murakami}
\end{equation}
where  $c_2(x)=\int^x_0 dt \big(\ln\frac{1+t}{t}\big)^2$ and $\Omega_{nk}=i\epsilon_{\mu\nu}\Bigg[ \sigma_3 \frac{\partial T^{\dagger}_{k}}{\partial k_{\mu}} \sigma_3 \frac{\partial T_{k}}{\partial k_{\nu}} \Bigg]$ is the Berry curvature of the $n^{\text{th}}$ boson band. $T_{k}$ is the Bloch eigenstate and $n_B$ is the Bose distribution function. This expression has been the most widely used in the Kitaev literature.

A general real space expression for thermal Hall response of arbitrary 2D Hamiltonians with local interactions was provided by Kitaev \cite{kitaev2006anyons} and subsequently clarified by Kapustin \cite{kapustin2020thermal}. A significant point made in these papers was that linear response theory does not directly yield $\kappa_{xy}$ but instead changes in $\kappa_{xy}/T$ between different temperatures, and that the correct temperature dependence of $\kappa_{xy}$ can be obtained once the value of $\kappa_{xy}/T$ is known for a particular temperature. 

Application of Kitaev-Kapustin theory for noninteracting bosons yields \cite{guo2020gauge} an expression for the derivative of $\kappa_{xy}/T:$

\begin{align}
    \frac{d(\kappa_{xy}/T)}{dT}=\frac{1}{2T^3 V}\sum_{k} \sum_n n'_{B}(\xi_{ik})\xi^3_{ik}\Omega^z_{nk}.
    \label{eq:guo}
\end{align}
The sum runs over both positive and negative eigenmodes with equal contribution. The thermal Hall conductivity is obtained by integrating Eq. (\ref{eq:guo}), and the constant of integration is fixed by specifying $\kappa_{xy}/T$ at some temperature. For instance, for a bounded spectrum, $\lim_{T\rightarrow\infty} (\kappa_{xy}/T)=0.$ 

For interacting bosons, one needs to revert again to Kitaev-Kapustin theory and obtain the thermal Hall response in terms of the interacting Green functions. To the best of our knowledge, this has not been done yet.

In an earlier work \cite{kumar2022absence}, two of us observed that $\kappa_{xy}$ for the Kitaev model numerically obtained from Eq. (\ref{eq:murakami}) for moderate fields $\mathbf{h} \sim K(111)$ shows a tendency to saturate \cite{mcclarty2018topological} to a finite value even at high temperatures. The same is the case when $\kappa_{xy}$ is obtained from Eq. (\ref{eq:guo}). This was puzzling on two counts. First, since the ``infinite'' temperature density matrix is diagonal with equal elements, the energy currents and $\kappa_{xy}$ should vanish. Second, physically, thermal response should be proportional to the specific heat which vanishes at high temperatures in any system with a bounded spectrum. In Ref. \cite{kumar2022absence} a new technique for the calculation of the thermal Hall response of gapped 2D spin systems was proposed which is simpler to implement numerically compared to the Kitaev-Kapustin linear response treatment, which was used to decide between Eq. (\ref{eq:murakami}) and Eq. (\ref{eq:guo}). The basic idea is that for any local Hamiltonian, i.e. $H = \sum_{i} H_{i},$ where $i$ refer to spatial coordinates, the thermal current $j_{E}(ij)$ from $i$ to $j$ is given by $j_{E}(ij)=i[H_i,H_j].$ The energy current across a surface is found by summing such local energy current contributions such that the two sites in each local contribution $j_{E}(ij)$ lie on either side of the boundary. For Hall response, the 2D lattice is rolled up into a cylinder, a time reversal symmetry breaking perturbation (here, a Zeeman field) is applied, and the system is immersed in a \textit{uniform} temperature bath. The thermal Hall current is measured across any simple curve (boundary) that connects the two open edges. Since both the edges are at the same temperature, the total Hall current is zero, a consequence of the exact cancellation of Hall currents reckoned from the two edges. For gapped systems, the Hall currents are the largest near the edges and exponentially decay into the bulk. Thus one obtains the Hall current associated with a single edge at any given temperature $T$ by computing the energy current across a curve that extends from one of the edges to the midpoint of a sufficiently large cylinder. For calculation of the thermal averages, a purification based tensor network technique was used. Thermal Hall response of the Kitaev model as well as Kitaev-Heisenberg models were thus obtained. The numerical strategy does not require any quasiparticle picture. 

\begin{figure}
\centering
\includegraphics[width=\columnwidth]{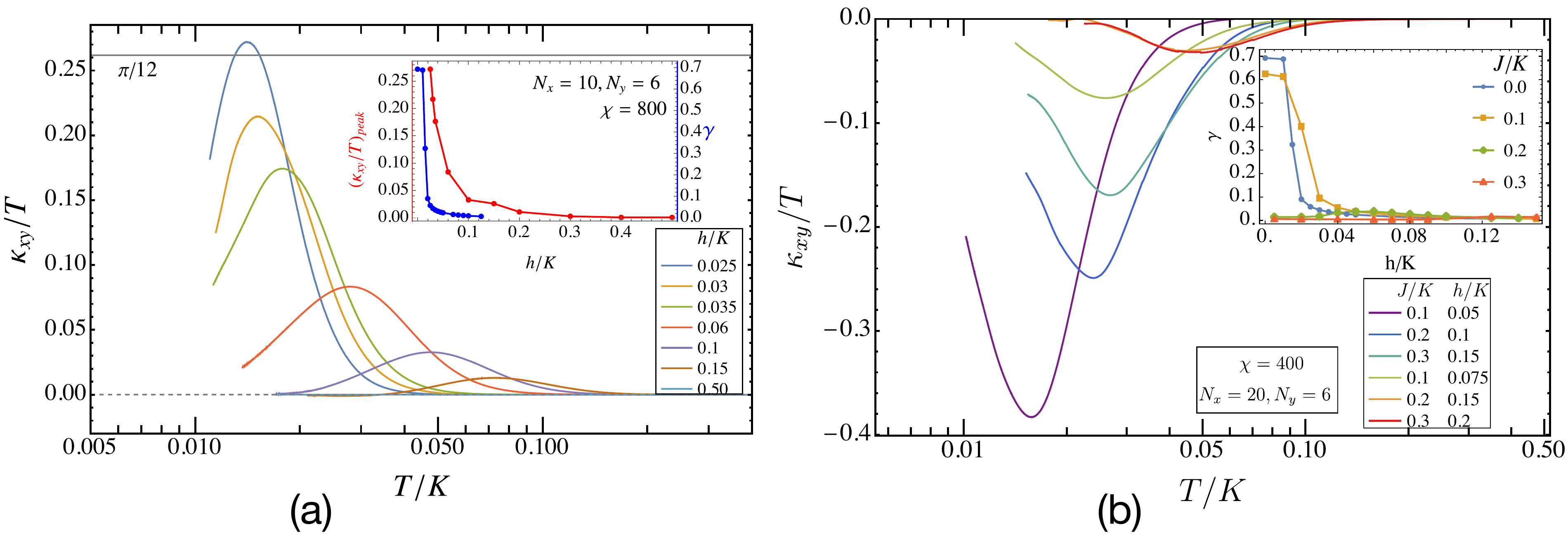}
\caption{\label{fig:thermal_hall} (a) (Reprinted with permission from Ref. \cite{kumar2022absence}) shows isomagnetic plots of $\kappa_{xy}/T$ for the ferromagnetic Kitaev model in the field range $0.025 \leq h/K \leq 0.5$ interpolating the Ising topological order (ITO) and spin wave limits. Solid gray line describes the half-quantized thermal Hall effect. Inset shows the plot of peaked $\kappa_{xy}/T$ versus magnetic field on the left $y$ axes and topological entanglement entropy $\gamma$ on the right $y$ axes in pure Kitaev limit. In (b) the same is shown for the $J-K$ model, with Heisenberg interactions chosen both inside the Kitaev spin liquid phase ($J/K = 0.1$) as well as in the SDW phase ($J/K = 0.2, 0.3$) whereas the inset shows the plot of topological entanglement entropy $\gamma$ as a function of magnetic field $h/K$ for different values of $J/K.$ }
\end{figure}

Figure \ref{fig:thermal_hall} shows the thermal Hall response of the Kitaev model in a $(111)$ Zeeman field calculated using the technique introduced in Ref. \cite{kumar2022absence}. Because of the finite size nature of the samples, the edge modes become gapped even at vanishing fields. At higher fields when ITO is lost, the excitation gaps are genuine and not a finite size effect. At the lowest fields, $\kappa_{xy}/T$ peaks near the half-quantized value expected for the ITO phase. Thereafter, the peak rapidly declines upon increasing the field beyond the ITO regime. At a field $\mathbf{h} = 0.5 K (111),$ the thermal Hall conductivity  is practically zero, unlike the results obtained from spin wave theory. No saturation of $\kappa_{xy}$ is observed at high temperatures. We believe that this difference is on account of the inapplicability of spin wave theory in such a field regime where nonlinear magnon interaction effects could be significant. The topological entanglement entropy (see inset) also shows a similar behaviour, declining sharply from $\gamma=\ln 2$ outside the ITO phase. At the higher range of fields far from ITO, the spin wave approximation becomes justifiable. 

When perturbed by Heisenberg or other competing spin interactions, the ground state becomes magnetically ordered. Thermal Hall effect has been studied in such cases using the spin wave approximation around the classical magnetic ground states. Ref. \cite{mcclarty2018topological} describes  that  the spin-wave bands of the $K-J-\Gamma-\Gamma'$  model carry nontrivial Chern numbers over large regions of the phase diagram, implying the presence of chiral magnon edge states. Unlike other models of topological magnons here  nonvanishing anomalous (number nonconserving) terms 
 are responsible for opening up a gap in the spectrum leading to Chern bands. The evidence of the chiral surface states that are present and topologically protected in linear spin-wave theory  is shown from the time-dependent DMRG  and interacting spin-wave theory, and hence, should be experimentally detectable in principle. These chiral magnons lead to integer quantized Hall response that reverses its  sign upon the reversal of the  magnetic field \cite{zhang2021topological}. A direct numerical calculation \cite{kumar2022absence} for the Kitaev-Heisenberg model in the vicinity of field suppressed magnetic order reveals a large finite temperature peak (approaching integer quantization for $J/K$ near the Kitaev spin liquid boundary), see Figure. However at low as well as high temperatures, the thermal Hall response approaches zero. This is reminiscent of recent observations \cite{czajka2021oscillations}. A calculation of the topological entanglement entropy \cite{kumar2022absence} does not indicate any restoration of ITO near field suppressed magnetic order. It appears that although the magnetic field is determental to magnetic ordering, in parallel it also acts to suppress topological order.
 
\textit{Phonon contribution to Hall effect}:
Although phonons are not intrinsically chiral and do not directly contribute to the thermal Hall effect, the phonons inevitably couple to the magnetic excitations in the Kitaev system, owing to the dependence of the spin interactions on the interatomic separations. The phonon contributions could have both intrinsic origin (from Berry phase effects of the hybrid magnon-phonon modes) as well as extrinsic ones associated with defect scattering (side jump and/or skew symmetric origin). The theory of phonon contribution to thermal Hall effect through their coupling to spin waves has been recently developed in Ref. \cite{joshi2018topological}. Using these ideas, it has been argued in Ref. \cite{li2023magnons} that to explain the experimentally observed thermal Hall response in the vicinity of field-suppressed magnetic order, one must include both intrinsic and extrinsic contributions to the thermal Hall effect. Being a spin-wave analysis, these studies however do not shed light on whether there is any signature of fractionalized excitations in this field regime. 

\textit{Other thermal response}: Experimentally, we have seen how the field dependence of the spectral gap extracted from specific heat measurements could provide convincing evidence for the existence of fractionalized Kitaev quasiparticles. We are not aware of theoretical work that supports this picture

\textit{Disorder as a signature of fractionalized excitations:} There have also been theoretical and experimental studies on the consequences of various types of disorder in quantum
spin liquid systems. Most materials realistically have some level of disorder, which can significantly affect their low-energy
properties, but can also serve as a probe for exotic excitations
characterizing spin liquid systems. In this context, more than
a decade ago, both site dilution and exchange randomness were considered
in the Kitaev honeycomb model, and it was shown that in the gapless
phase, a single vacancy binds a flux and induces a local moment,
leading to a logarithmically divergent local susceptibility \cite{PhysRevLett.104.237203,das2016kondo}. This effect is further enhanced for a pair of nearby vacancies.
On the other hand, in the gapped phase, it was shown that a finite
density of randomly distributed vacancies remains tractable via a
mapping to a bipartite random hopping problem \cite{PhysRevB.84.115146}, which leads to a strong disorder form of the low-energy
thermodynamics. The effect
of coupling magnetic impurities to the honeycomb lattice spin-1/2
Kitaev model in its spin liquid phase also gives rise to an unusual Kondo effect \cite{PhysRevLett.105.117201} whose strong coupling limit creates localized nonabelian anyons just like in the vacancy case \cite{PhysRevLett.105.117201,das2016kondo}. Furthermore, the massless
spinons in the spin liquid were shown to mediate an interaction between distinct
impurities unlike the usual dipolar RKKY interaction noted in various
2D impurity problems.

More recently, other possible signatures of different types of disorder
in the Kitaev model have been considered, with a focus on intercalated
compounds, with weaker interlayer couplings, an effectively quasi-2D
nature and a large number of stacking faults, where defects are expected
to play an important role, and which are also relatively better described
by a pure Kitaev model without competing interactions. For example,
in the the hydrogen intercalated iridate H$_{3}$LiIr$_{2}$O$_{6}$,
vacancies in the Kitaev model have been found to play an important
role in understanding the origin of a low-temperature divergence in
the specific heat $C/T\propto T^{-1/2}$ as resulting from a pileup
of the low-energy density of states \cite{PhysRevLett.122.047202,PhysRevX.11.011034}. The vacancy-induced low-energy states
are predicted to be gapped out in the presence of a magnetic field,
resulting in the suppression of the specific heat, as observed
in this material. It was also predicted that Raman or inelastic neutron
scattering experiments should be able to detect the low-energy divergent
power-law tail of the density of states, and a controlled, sufficiently
large change of disorder concentration should be observable as a drift
in the exponent of this divergence. While such an interpretation is
indicative of the potentially useful role played by disorder in detecting
spin liquid physics, further discussion on other possible ways of understanding these experimental observations, unrelated to fractionalization
or topological order, is desirable. Besides, spin vacancies in the
non-abelian Kitaev spin liquid are known to be harbor Majorana zero
modes, the spectroscopic signatures of which have recently been studied
in a scanning tunneling setup \cite{kao2023vacancy} where a non-abelian
Kitaev spin liquid with a finite density of spin vacancies forms a
tunneling barrier between a tip and a substrate, and the key results
include a well-defined peak close to zero bias voltage in the derivative
of the tunneling conductance whose intensity increases with the density
of vacancies, and a single-fermion Van Hove singularity at a higher
voltage that reveals the scale of the emergent Majorana fermions in
the Kitaev spin liquid. The dynamical response of vacancy-induced
quasiparticle excitations in the site-diluted Kitaev spin liquid has
also been studied in the presence of a magnetic field \cite{kao2023dynamics}.
Due to the flux-binding effect and the emergence of dangling Majorana
fermions around each spin vacancy, the low-energy physics is governed
by a set of vacancy-induced quasi-zero-energy modes, resulting in
unique characteristics of the dynamical spin correlation functions.
\section{\label{sec:discuss} Discussion}
To conclude, we have surveyed the current theoretical and experimental understanding of Zeeman field tuned degradation of topological order and spin fractionalization in Kitaev magnets. Along with external  Zeeman field, effects of other competing perturbations like Heisenberg and Gamma interactions were also discussed. 

Among the signatures of fractionalization, data from inelastic neutron scattering measurements \cite{banerjee2016proximate,banerjee2018excitations,balz2019finite,zhao2022neutron,winter2017breakdown,winter2016challenges}, while not inconsistent with fractionalized excitations, can also be explained from an interacting spin waves perspective. A definitive confirmation of fractionalization is preferred over arguments merely suggesting compatibility with it. NMR studies \cite{janvsa2018observation} have demonstrated that the field-dependent spin excitation gap can be modeled by combining a finite value at zero field with a cubic increase in the low-field range. Given that a solitary spin flip is anticipated to yield a pair of visons and a Majorana in the Kitaev model, this behaviour in the low-field regime serves as positive  evidence for the presence of fractionalized Kitaev excitations. The cubic  behaviour can also be extracted from recent specific heat measurements a field configuration that has nonzero components along all three spin projection axes. NMR and specific heat measurements \cite{janvsa2018observation,tanaka2022thermodynamic}, in our opinion, represent the strongest evidence so far for fractionalization in field-tuned Kitaev magnets, although one must remember here that the cubic field dependence is for the Kitaev model, and it is not known if the same behaviour would persist in the presence of competing interactions strong enough to cause magnetic ordering. 

The most important debate regarding thermal measurements \cite{kasahara2018majorana, yokoi2021half, kasahara2022quantized,bruin2022robustness} concerns the controversial observation of a half-quantized thermal Hall conductivity near the field suppressed  magnetic order, where there is strong disagreement in the community even on the existence of the phenomenon in $\alpha$-RuCl$_3$ \cite{lefranccois2023oscillations,czajka2021oscillations}. The experiments are conducted in a field regime where neither spinons nor magnons provide a good quasiparticle starting point for analyzing thermal Hall response. To add to the confusion, it appears that spin-dependent scattering of phonons from electronic excitations and defects could strongly affect the thermal Hall response of Kitaev magnets. However there has been progress in the development of quasiparticle agnostic numerical approaches for the study of thermal Hall response in many-body systems \cite{kapustin2020thermal,guo2020gauge,kumar2022absence}. Second generation Kitaev materials, while heavily disordered, are less susceptible to magnetic order, and the known properties of defects in Kitaev spin liquids have been used to interpret specific heat data of second generation Kitaev materials as evidence of spin fractionalization \cite{PhysRevLett.122.047202,PhysRevX.11.011034}. The defect studies are, however, based on mean-field fermionization of the Kitaev model and do not take into account the gauge fluctuation contributions. 

Given the complicated nature of the problem when both Zeeman perturbations as well as competing spin interactions are simultaneously present, a significant theoretical focus has been directed towards the simpler problem of Zeeman effects in Kitaev systems. A small Zeeman field (with nonzero field components along the three natural spin quantization axes) gaps out the Majoranas and makes them chiral \cite{kitaev2006anyons}. In the case of a ferromagnetic (FM) sign for the Kitaev coupling, the system undergoes a direct transition into a partially spin-polarized phase, which is topologically trivial. The shift towards the topologically trivial phase seems to follow a confinement-deconfinement mechanism driven by the proliferation of visons. Conversely, in the antiferromagnetic (AFM) scenario \cite{zhu2018robust,gohlke2018dynamical,das2024microscopic,nasu2018successive}, there exists at least one intermediate field phase that separates the low-field intermediate topological order (ITO) phase from the high-field partially spin-polarized paramagnetic phase. In recent studies,  the intermediate field phase is shown to be abelian \cite{patel2019magnetic,jiang2018field,hickey2019emergence,hickey2021generic,zhang2022theory,das2024microscopic} and exhibits half quantized thermal Hall response \cite{nasu2018successive,das2024microscopic} indicating the half quantization is  a necessary but not sufficient condition for proving the Ising topological order, which additionally requires the demonstration of appropriate braiding statistics and fusion rules \cite{das2024microscopic}.

Competing spin interactions are necessary to account for various experimental observations, such as the existence of low-field antiferromagnetic zigzag order in $\alpha-$RuCl$_{3}.$  One possibility is that the Zeeman and competing spin interactions operate through distinct and competing mechanisms to disrupt topological order. This provides a motivation to employ both simultaneously, preventing either from fully suppressing the nearby topological order within the parameter space, while effectively counteracting their individual influence. Different numerical and analytical studies \cite{liu2018dirac,lee2020magnetic,li2021identification,yu2023nematic,zou2020field} have been developed in this context, focusing on fractionalization and topological order, but have not reported any evidence for field revived ITO, although there is some support for intermediate paramagnetic phases that could be regarded as abelian spin liquids. Moreover, in analytical investigations, the influence of gauge effects has typically been disregarded, impacting the reliability with which assertions can be made about the characteristics of topological order and fractionalized excitations.

Field-tuned strategies for Ising topological order in Kitaev systems could work in a few scenarios. First, based on the observation that models with the AFM sign of Kitaev interaction are significantly more robust to field induced ITO degradation, it makes sense to search for highly frustrated magnetic insulators with AFM Kitaev interactions. Second, even in FM Kitaev materials (which are more common so far), if the magnetic ordering scale happens to be well below the vison gap, then such systems can also be field-tuned to an ITO phase. Further work is also needed to understand better the nature of any paramagnetic phases that are obtained upon the Zeeman suppression of magnetic order in Kitaev models with competing spin interactions. 

The bulk of the theoretical treatment of Zeeman-perturbed Kitaev systems is based on quasiparticle mean field treatments, whether magnons or spinons. Effects of magnon interactions have been studied in the context of inelastic neutron scattering, although not on thermal Hall response. However, more work needs to be done to understand magnon interaction effects closer to spin liquid instabilities. From the spinon end, interactions are naturally introduced through coupling to suitable gauge fluctuations. Gauge theory approaches are still in the process of development and offer a complementary treatment to spin wave approaches. The effect of gauge fluctuations on physical properties such as thermal Hall response is a topic that is relevant for a wide range of 2D quantum magnets, including the cuprates.
\section*{Acknowledgements}
The authors acknowledge support of the Department of Atomic Energy, Government of India, under Project Identification No. RTI 4002, and Department of Theoretical Physics, TIFR, for computational resources. S.K. acknowledges financial support from UF Project No. P0224175 - Dirac postdoc fellowship, sponsored by the Florida State University National High Magnetic Field Laboratory (NHMFL), and from NSF DMR-2128556. A.K. thanks the National High Magnetic Field Laboratory, which is supported by National Science Foundation Cooperative Agreement No. DMR-2128556* and the State of Florida. A.K. was supported through a Dirac postdoctoral fellowship at NHMFL. The  authors  thank  Subir Sachdev, Shiraz Minwalla, Kedar Damle, Darshan Joshi, Avijit Maity, Inti Sodemann, Nandini Trivedi, Wen-Han Kao, Yong-Baek Kim and Matthias Gohlke for fruitful  discussions. 
\section*{References}
\bibliographystyle{iopart-num}
\bibliography{References}

\providecommand{\newblock}{}
\begin{thebibliography}{10}
\expandafter\ifx\csname url\endcsname\relax
  \def\url#1{{\tt #1}}\fi
\expandafter\ifx\csname urlprefix\endcsname\relax\def\urlprefix{URL }\fi
\providecommand{\eprint}[2][]{\url{#2}}

\bibitem{anderson1987resonating}
Anderson P~W 1987 {\em Science\/} {\bf 235} 1196--1198

\bibitem{baskaran1993resonating}
Baskaran G, Zou Z and Anderson P~W 1993 {\em Solid state commun.\/} {\bf 88} 853--856

\bibitem{kalmeyer1987equivalence}
Kalmeyer V and Laughlin R 1987 {\em Phys. Rev. Lett.\/} {\bf 59} 2095

\bibitem{kivelson1987topology}
Kivelson S~A, Rokhsar D~S and Sethna J~P 1987 {\em Phys. Rev. B\/} {\bf 35} 8865

\bibitem{affleck1988large}
Affleck I and Marston J~B 1988 {\em Phys. Rev. B\/} {\bf 37} 3774

\bibitem{read1991large}
Read N and Sachdev S 1991 {\em Phys. Rev. Lett.\/} {\bf 66} 1773

\bibitem{wen1989vacuum}
Wen X~G 1989 {\em Phys. Rev. B\/} {\bf 40} 7387

\bibitem{wegner1971duality}
Wegner F~J 1971 {\em J. of Math. Phys.\/} {\bf 12} 2259--2272

\bibitem{senthil2000z}
Senthil T and Fisher M~P 2000 {\em Phys. Rev. B\/} {\bf 62} 7850

\bibitem{kogut1979introduction}
Kogut J~B 1979 {\em Rev. Mod. Phys.\/} {\bf 51} 659

\bibitem{moessner2001resonating}
Moessner R and Sondhi S 2001 {\em Phys. Rev. Lett.\/} {\bf 86} 1881

\bibitem{kitaev2006anyons}
Kitaev A 2006 {\em Ann. Phys. (N. Y.)\/} {\bf 321} 2--111

\bibitem{savary2016quantum}
Savary L and Balents L 2016 {\em Rep. on Prog. in Phys.\/} {\bf 80} 016502

\bibitem{ivanov2001non}
Ivanov D~A 2001 {\em Phys. Rev. Lett.\/} {\bf 86} 268

\bibitem{janssen2019heisenberg}
Janssen L and Vojta M 2019 {\em Journal of Physics: condensed matter\/} {\bf 31} 423002

\bibitem{hermanns2018physics}
Hermanns M, Kimchi I and Knolle J 2018 {\em Annual Review of Condensed Matter Physics\/} {\bf 9} 17--33

\bibitem{takagi2019concept}
Takagi H, Takayama T, Jackeli G, Khaliullin G and Nagler S~E 2019 {\em Nature Reviews Physics\/} {\bf 1} 264--280

\bibitem{trebst2022kitaev}
Trebst S and Hickey C 2022 {\em Physics Reports\/} {\bf 950} 1--37

\bibitem{kumar2020kitaev}
Kumar A and Tripathi V 2020 {\em Physical Review B\/} {\bf 102} 100401

\bibitem{banerjee2016proximate}
Banerjee A, Bridges C, Yan J~Q, Aczel A, Li L, Stone M, Granroth G, Lumsden M, Yiu Y, Knolle J {\em et~al.\/} 2016 {\em Nature materials\/} {\bf 15} 733--740

\bibitem{winter2016challenges}
Winter S~M, Li Y, Jeschke H~O and Valent{\'\i} R 2016 {\em Physical Review B\/} {\bf 93} 214431

\bibitem{kasahara2018majorana}
Kasahara Y, Ohnishi T, Mizukami Y, Tanaka O, Ma S, Sugii K, Kurita N, Tanaka H, Nasu J, Motome Y {\em et~al.\/} 2018 {\em Nature\/} {\bf 559} 227--231

\bibitem{janvsa2018observation}
Jan{\v{s}}a N, Zorko A, Gomil{\v{s}}ek M, Pregelj M, Kr{\"a}mer K~W, Biner D, Biffin A, R{\"u}egg C and Klanj{\v{s}}ek M 2018 {\em Nature physics\/} {\bf 14} 786--790

\bibitem{banerjee2018excitations}
Banerjee A, Lampen-Kelley P, Knolle J, Balz C, Aczel A~A, Winn B, Liu Y, Pajerowski D, Yan J, Bridges C~A {\em et~al.\/} 2018 {\em npj Quantum Materials\/} {\bf 3} 8

\bibitem{ran2017spin}
Ran K, Wang J, Wang W, Dong Z~Y, Ren X, Bao S, Li S, Ma Z, Gan Y, Zhang Y {\em et~al.\/} 2017 {\em Physical review letters\/} {\bf 118} 107203

\bibitem{wang2017theoretical}
Wang W, Dong Z~Y, Yu S~L and Li J~X 2017 {\em Physical Review B\/} {\bf 96} 115103

\bibitem{zhao2022neutron}
Zhao X, Ran K, Wang J, Bao S, Shangguan Y, Huang Z, Liao J, Zhang B, Cheng S, Xu H {\em et~al.\/} 2022 {\em Chinese Physics Letters\/} {\bf 39} 057501

\bibitem{balz2019finite}
Balz C, Lampen-Kelley P, Banerjee A, Yan J, Lu Z, Hu X, Yadav S~M, Takano Y, Liu Y, Tennant D~A {\em et~al.\/} 2019 {\em Physical Review B\/} {\bf 100} 060405

\bibitem{winter2017breakdown}
Winter S~M, Riedl K, Maksimov P~A, Chernyshev A~L, Honecker A and Valent{\'\i} R 2017 {\em Nature communications\/} {\bf 8} 1152

\bibitem{wulferding2020magnon}
Wulferding D, Choi Y, Do S~H, Lee C~H, Lemmens P, Faugeras C, Gallais Y and Choi K~Y 2020 {\em Nature communications\/} {\bf 11} 1603

\bibitem{sahasrabudhe2023chiral}
Sahasrabudhe A, Prosnikov M~A, Koethe T~C, Stein P, Tsurkan V, Loidl A, Gr{\"u}ninger M, Hedayat H and van Loosdrecht P~H 2023 {\em arXiv preprint arXiv:2305.03400\/}

\bibitem{wu2018field}
Wu L, Little A, Aldape E~E, Rees D, Thewalt E, Lampen-Kelley P, Banerjee A, Bridges C~A, Yan J~Q, Boone D {\em et~al.\/} 2018 {\em Physical Review B\/} {\bf 98} 094425

\bibitem{nagai2020two}
Nagai Y, Jinno T, Yoshitake J, Nasu J, Motome Y, Itoh M and Shimizu Y 2020 {\em Physical Review B\/} {\bf 101} 020414

\bibitem{yamashita2010highly}
Yamashita M, Nakata N, Senshu Y, Nagata M, Yamamoto H~M, Kato R, Shibauchi T and Matsuda Y 2010 {\em Science\/} {\bf 328} 1246--1248

\bibitem{yokoi2021half}
Yokoi T, Ma S, Kasahara Y, Kasahara S, Shibauchi T, Kurita N, Tanaka H, Nasu J, Motome Y, Hickey C {\em et~al.\/} 2021 {\em Science\/} {\bf 373} 568--572

\bibitem{kasahara2022quantized}
Kasahara Y, Suetsugu S, Asaba T, Kasahara S, Shibauchi T, Kurita N, Tanaka H and Matsuda Y 2022 {\em Physical Review B\/} {\bf 106} L060410

\bibitem{bruin2022robustness}
Bruin J, Claus R, Matsumoto Y, Kurita N, Tanaka H and Takagi H 2022 {\em Nat. Phys.\/} {\bf 18} 401--405

\bibitem{lefranccois2023oscillations}
Lefran{\c{c}}ois {\'E}, Baglo J, Barth{\'e}lemy Q, Kim S, Kim Y~J and Taillefer L 2023 {\em Physical Review B\/} {\bf 107} 064408

\bibitem{czajka2021oscillations}
Czajka P, Gao T, Hirschberger M, Lampen-Kelley P, Banerjee A, Yan J, Mandrus D~G, Nagler S~E and Ong N 2021 {\em Nature Physics\/} {\bf 17} 915--919

\bibitem{tanaka2022thermodynamic}
Tanaka O, Mizukami Y, Harasawa R, Hashimoto K, Hwang K, Kurita N, Tanaka H, Fujimoto S, Matsuda Y, Moon E~G {\em et~al.\/} 2022 {\em Nature Physics\/} {\bf 18} 429--435

\bibitem{widmann2019thermodynamic}
Widmann S, Tsurkan V, Prishchenko D~A, Mazurenko V~G, Tsirlin A~A and Loidl A 2019 {\em Physical Review B\/} {\bf 99} 094415

\bibitem{schonemann2020thermal}
Sch{\"o}nemann R, Imajo S, Weickert F, Yan J, Mandrus D~G, Takano Y, Brosha E~L, Rosa P~F, Nagler S~E, Kindo K {\em et~al.\/} 2020 {\em Physical Review B\/} {\bf 102} 214432

\bibitem{das2019magnetic}
Das S~D, Kundu S, Zhu Z, Mun E, McDonald R~D, Li G, Balicas L, McCollam A, Cao G, Rau J~G {\em et~al.\/} 2019 {\em Physical Review B\/} {\bf 99} 081101

\bibitem{leahy2017anomalous}
Leahy I~A, Pocs C~A, Siegfried P~E, Graf D, Do S~H, Choi K~Y, Normand B and Lee M 2017 {\em Physical review letters\/} {\bf 118} 187203

\bibitem{wellm2018signatures}
Wellm C, Zeisner J, Alfonsov A, Wolter A, Roslova M, Isaeva A, Doert T, Vojta M, B{\"u}chner B and Kataev V 2018 {\em Physical Review B\/} {\bf 98} 184408

\bibitem{johnson2015monoclinic}
Johnson R~D, Williams S, Haghighirad A, Singleton J, Zapf V, Manuel P, Mazin I, Li Y, Jeschke H~O, Valent{\'\i} R {\em et~al.\/} 2015 {\em Physical Review B\/} {\bf 92} 235119

\bibitem{kubota2015successive}
Kubota Y, Tanaka H, Ono T, Narumi Y and Kindo K 2015 {\em Physical Review B\/} {\bf 91} 094422

\bibitem{bachus2020thermodynamic}
Bachus S, Kaib D~A, Tokiwa Y, Jesche A, Tsurkan V, Loidl A, Winter S~M, Tsirlin A~A, Valenti R and Gegenwart P 2020 {\em Physical Review Letters\/} {\bf 125} 097203

\bibitem{zhou2023possible}
Zhou X~G, Li H, Matsuda Y~H, Matsuo A, Li W, Kurita N, Su G, Kindo K and Tanaka H 2023 {\em Nature Communications\/} {\bf 14} 5613

\bibitem{zhu2018robust}
Zhu Z, Kimchi I, Sheng D and Fu L 2018 {\em Phys. Rev. B\/} {\bf 97} 241110

\bibitem{das2024microscopic}
Das J and Tripathi V 2024 {\em arXiv preprint arXiv:2401.12750\/}

\bibitem{gohlke2018dynamical}
Gohlke M, Moessner R and Pollmann F 2018 {\em Phys. Rev. B\/} {\bf 98} 014418

\bibitem{joy2022dynamics}
Joy A~P and Rosch A 2022 {\em Phys. Rev. X\/} {\bf 12} 041004

\bibitem{chen2023nature}
Chen C and Villadiego I~S 2023 {\em Phys. Rev. B\/} {\bf 107} 045114

\bibitem{patel2019magnetic}
Patel N~D and Trivedi N 2019 {\em Proceedings of the National Academy of Sciences\/} {\bf 116} 12199--12203

\bibitem{jiang2018field}
Jiang H~C, Wang C~Y, Huang B and Lu Y~M 2018 {\em arXiv preprint arXiv:1809.08247\/}

\bibitem{hickey2019emergence}
Hickey C and Trebst S 2019 {\em Nature communications\/} {\bf 10} 530

\bibitem{hickey2021generic}
Hickey C, Gohlke M, Berke C and Trebst S 2021 {\em Physical Review B\/} {\bf 103} 064417

\bibitem{zhang2022theory}
Zhang S~S, Hal{\'a}sz G~B and Batista C~D 2022 {\em Nat. Commun.\/} {\bf 13} 399

\bibitem{jiang2020tuning}
Jiang M~H, Liang S, Chen W, Qi Y, Li J~X and Wang Q~H 2020 {\em Physical Review Letters\/} {\bf 125} 177203

\bibitem{yogendra2023emergent}
Yogendra K, Das T and Baskaran G 2023 {\em Physical Review B\/} {\bf 108} 165118

\bibitem{schellenberger2022dynamic}
Schellenberger A, H{\"o}rmann M and Schmidt K~P 2022 {\em Physical Review B\/} {\bf 106} 104403

\bibitem{nasu2018successive}
Nasu J, Kato Y, Kamiya Y and Motome Y 2018 {\em Phys. Rev. B\/} {\bf 98} 060416

\bibitem{berke2020field}
Berke C, Trebst S and Hickey C 2020 {\em Physical Review B\/} {\bf 101} 214442

\bibitem{feng2023dimensional}
Feng S, Agarwala A and Trivedi N 2023 {\em arXiv preprint arXiv:2308.08116\/}

\bibitem{das2023jordan}
Das J, Kumar A, Maity A and Tripathi V 2023 {\em Phys. Rev. B\/} {\bf 108} 085110

\bibitem{samajdar2019enhanced}
Samajdar R, Scheurer M~S, Chatterjee S, Guo H, Xu C and Sachdev S 2019 {\em Nature Physics\/} {\bf 15} 1290--1294

\bibitem{liu2018dirac}
Liu Z~X and Normand B 2018 {\em Physical review letters\/} {\bf 120} 187201

\bibitem{lee2020magnetic}
Lee H~Y, Kaneko R, Chern L~E, Okubo T, Yamaji Y, Kawashima N and Kim Y~B 2020 {\em Nature communications\/} {\bf 11} 1639

\bibitem{li2021identification}
Li H, Zhang H~K, Wang J, Wu H~Q, Gao Y, Qu D~W, Liu Z~X, Gong S~S and Li W 2021 {\em Nature Communications\/} {\bf 12} 4007

\bibitem{yu2023nematic}
Yu S~Y, Li H, Zhao Q~R, Gao Y, Dong X~Y, Liu Z~X, Li W and Gong S~S 2023 {\em arXiv preprint arXiv:2304.00555\/}

\bibitem{zou2020field}
Zou L and He Y~C 2020 {\em Phys. Rev. Res.\/} {\bf 2} 013072

\bibitem{nasu2019nonequilibrium}
Nasu J and Motome Y 2019 {\em Physical Review Research\/} {\bf 1} 033007

\bibitem{harada2023field}
Harada C, Ono A and Nasu J 2023 {\em arXiv preprint arXiv:2305.08357\/}

\bibitem{thouless1982quantized}
Thouless D~J, Kohmoto M, Nightingale M~P and den Nijs M 1982 {\em Physical review letters\/} {\bf 49} 405

\bibitem{cooper1997thermoelectric}
Cooper N, Halperin B and Ruzin I 1997 {\em Physical Review B\/} {\bf 55} 2344

\bibitem{qin2011energy}
Qin T, Niu Q and Shi J 2011 {\em Physical review letters\/} {\bf 107} 236601

\bibitem{matsumoto2014thermal}
Matsumoto R, Shindou R and Murakami S 2014 {\em Physical Review B\/} {\bf 89} 054420

\bibitem{guo2020gauge}
Guo H, Samajdar R, Scheurer M~S and Sachdev S 2020 {\em Physical Review B\/} {\bf 101} 195126

\bibitem{kapustin2020thermal}
Kapustin A and Spodyneiko L 2020 {\em Physical Review B\/} {\bf 101} 045137

\bibitem{kumar2022absence}
Kumar A and Tripathi V 2023 {\em Phys. Rev. B\/} {\bf 107} L220406

\bibitem{mcclarty2018topological}
McClarty P, Dong X~Y, Gohlke M, Rau J, Pollmann F, Moessner R and Penc K 2018 {\em Physical Review B\/} {\bf 98} 060404

\bibitem{zhang2021topological}
Zhang E~Z, Chern L~E and Kim Y~B 2021 {\em Physical Review B\/} {\bf 103} 174402

\bibitem{joshi2018topological}
Joshi D~G 2018 {\em Phys. Rev. B\/} {\bf 98} 060405

\bibitem{li2023magnons}
Li S, Yan H and Nevidomskyy A~H 2023 {\em arXiv preprint arXiv:2301.07401\/}

\bibitem{PhysRevLett.104.237203}
Willans A~J, Chalker J~T and Moessner R 2010 {\em Phys. Rev. Lett.\/} {\bf 104}(23) 237203

\bibitem{das2016kondo}
Das S, Dhochak K and Tripathi V 2016 {\em Physical Review B\/} {\bf 94} 024411

\bibitem{PhysRevB.84.115146}
Willans A~J, Chalker J~T and Moessner R 2011 {\em Phys. Rev. B\/} {\bf 84}(11) 115146

\bibitem{PhysRevLett.105.117201}
Dhochak K, Shankar R and Tripathi V 2010 {\em Phys. Rev. Lett.\/} {\bf 105}(11) 117201

\bibitem{PhysRevLett.122.047202}
Knolle J, Moessner R and Perkins N~B 2019 {\em Phys. Rev. Lett.\/} {\bf 122}(4) 047202

\bibitem{PhysRevX.11.011034}
Kao W~H, Knolle J, Hal\'asz G~B, Moessner R and Perkins N~B 2021 {\em Phys. Rev. X\/} {\bf 11}(1) 011034

\bibitem{kao2023vacancy}
Kao W~H, Perkins N~B and Hal\'asz G~B 2023 {\em arXiv preprint arXiv:2307.10376\/}

\bibitem{kao2023dynamics}
Kao W~H, Hal\'asz G~B and Perkins N~B 2023 {\em arXiv preprint arXiv:2310.06891\/}

\end{thebibliography}

\end{document}